\documentclass[times]{aastex62}

\usepackage{CJK}
\usepackage{enumitem}
\usepackage{amsmath}
\usepackage{multirow}
\usepackage{cases}
\usepackage{graphicx}
\usepackage{subfigure}
\usepackage{natbib}
\usepackage{color}
\usepackage{tabularx}
\usepackage{bm}
\usepackage{threeparttable}
\usepackage{gensymb}
\usepackage{longtable}

\newcommand{\HL}[1]{\textcolor{black}{#1}}

\newcommand*{\dd}{\mathop{}\!\mathrm{d}}
\newcommand{\tE}{t_{\rm E}}

\newcommand{\e}{\varepsilon}
\newcommand{\eanom}{\varepsilon_{\rm anom}}
\newcommand{\edeg}{\HL{\varepsilon_{\rm uniq}}}

\DeclareGraphicsExtensions{.pdf,.png,.jpg}

\shorttitle{Binary-Lens Degeneracy \& Sensitivity}
\shortauthors{Shang et al.}

\begin{document}
\begin{CJK*}{UTF8}{gbsn}
\title{{\large Binary-lens Microlensing Degeneracy: Impact on Planetary Sensitivity and Mass-ratio Function}}

\correspondingauthor{Hongjing Yang}
\email{hongjing.yang@qq.com}

\author[0009-0005-0410-8451]{Yuxin Shang (尚钰欣)}
\affiliation{Department of Astronomy, Tsinghua University, Beijing 100084, China}

\author[0000-0003-0626-8465]{Hongjing Yang (杨弘靖)}
\affiliation{Department of Astronomy, School of Science, Westlake University, Hangzhou, Zhejiang 310030, China}

\author[0000-0002-1279-0666]{Jiyuan Zhang (张纪元)}
\affiliation{Department of Astronomy, Tsinghua University, Beijing 100084, China}

\author[0000-0001-8317-2788]{Shude Mao (毛淑德)}
\affiliation{Department of Astronomy, School of Science, Westlake University, Hangzhou, Zhejiang 310030, China}

\author{Andrew Gould} 
\affiliation{Department of Astronomy, Ohio State University, 140 W. 18th Ave., Columbus, OH 43210, USA}

\author[0000-0001-6000-3463]{Weicheng Zang (臧伟呈)}
\affiliation{Center for Astrophysics $|$ Harvard \& Smithsonian, 60 Garden St.,Cambridge, MA 02138, USA}

\author[0000-0003-4625-8595]{Qiyue Qian} 
\affiliation{Department of Astronomy, Tsinghua University, Beijing 100084, China}

\author[0000-0001-9481-7123]{Jennifer C. Yee}
\affiliation{Center for Astrophysics $|$ Harvard \& Smithsonian, 60 Garden St.,Cambridge, MA 02138, USA}

\begin{abstract}

Gravitational microlensing is a unique method for discovering cold planets across a broad mass range. Reliable statistics of the microlensing planets require accurate sensitivity estimates. However, the impact of the degeneracies in binary-lens single-source (2L1S) models that affect many actual planet detections is often omitted in sensitivity estimates, leading to potential self-inconsistency of the statistics studies.
In this work, we evaluate the effect of the 2L1S degeneracies on planetary sensitivity by simulating a series of typical microlensing events and comprehensively replicating \HL{a} realistic planet detection pipeline, including the anomaly identification, global 2L1S model search, and degenerate model comparison.
We find that \HL{for a pure-survey statistical sample, the} 2L1S degeneracies reduce the overall planetary sensitivity by $5-10\%$, with the effect increasing at higher planet-host mass ratios. This bias leads to an underestimation of planet occurrence rates and a flattening of the inferred mass-ratio function slope. This effect will be critical for upcoming space-based microlensing surveys like the Roman or Earth 2.0 missions, which are expected to discover $\mathcal{O}(10^3)$ planets. We also discuss the computational challenges and propose potential approaches for future applications.

\end{abstract}

\section{Introduction}\label{sec:intro}

The gravitational microlensing technique has enabled the detection of more than 240 exoplanets\footnote{\url{exoplanetarchive.ipac.caltech.edu}, as of June 2025} \citep{MaoPaczyski1991, GouldLoeb1992}. As a unique method sensitive to cold planets across a broad mass range, microlensing provides valuable complementary statistical properties of planets compared to other detection methods. Pioneering work by \citet{Gould2010_mufun_stat} and \citet{Cassan2012Nature} established that cold planets are common around microlensing hosts. Subsequent studies by \citet{Suzuki2016}, \citet{ShvartzvaldYossi2016_Wise}, \citet{Poleski2021_WideOrbitStatistic}, and \citet{Zang2025Science_KMT_MassRatioFunction} have presented detailed planet-host mass-ratio functions for microlensing planets. Although limited by small-sample statistics, these studies show some discrepancies in the precise form of the mass-ratio function; nonetheless, they consistently report an occurrence rate of cold planets of $\sim\mathcal{O}(1)$ per star.

Accurate measurement of the mass-ratio function requires a precise estimation of the sample sensitivity to planets as a function of mass ratio. The sample sensitivity, also referred to as the detection efficiency, is defined as the sum of the sensitivities of all microlensing events in the statistical sample. By definition, the sensitivity of an individual event represents the probability that a planet with given parameters would be detected and correctly characterized under the actual observational conditions, including data sampling and signal-to-noise ratio \HL{(SNR)} \citep[e.g.,][]{Rhie2000_MACHO9835_Sensitivity, Dong2006_OB04343_Sensitivity}. Ideally, the sensitivity calculation pipeline should replicate all steps involved in the actual planet detection process.

The general procedure for planetary detection in a microlensing light curve is as follows\footnote{For simplicity, we assume the lens system contains at most one host and one planet; in reality, more planets or stellar companions may be present.}:
\begin{enumerate}
    \item Fit the light curve with a \HL{single-lens, single-source (1L1S)} model \citep{Paczynski1986, WittMao1994_FiniteSource}, and then search for residual signals (anomalies) that may indicate the presence of a secondary lens, such as a planet. Anomalies are typically identified using $\Delta\chi^2$ thresholds \HL{or visual inspections} \citep[e.g.,][]{Yee2013_MB10311, Poleski2021_WideOrbitStatistic, Zang2021_ob191053_AnomalyFinder1, Yang2025_RAMP2}. If an anomaly is detected, the light curve proceeds to the next step. 
    
    \item Fit the light curve with \HL{binary-lens single-source} (2L1S) models. This involves a global exploration of the full parameter space to identify all plausible models. These candidate models are then optimized and compared. Again, a $\Delta\chi^2$ threshold is typically applied. If the viable models suggest a planetary mass ratio, the light curve proceeds to the next step.

    \item Examine whether the light curve can be explained by other physical effects. For example, the microlensing parallax effect \citep{Gould1992,Gould2000_Formalism,Gouldpies2004}, the presence of an additional source star \citep{Gaudi1998_1L2S}, or the xallarap effect \citep{Griest1992_1L2SEffect, Han_Gould_1997_1L2Sxarallap}. If none of these alternative interpretations can invalidate the planetary model, the planet is considered a robust detection.
\end{enumerate}

The first step, ``signal detection'', can be clearly defined using a $\Delta\chi^2$ threshold and has been widely adopted in previous sensitivity calculations. In contrast, the subsequent ``characterization'' steps are more complex, particularly the second step involving the parameter search and model selection, and have not been fully accounted for in earlier studies. This discrepancy between the actual detection process and the sensitivity calculation may introduce biases in the final sensitivity estimates.

The necessity of a global search arises from the severe degeneracy of the 2L1S microlensing light curves, resulting from the fact that the light curve is only a one-dimensional sample of the magnification map on the lens/source plane. Fortunately, in many cases, degenerate models yield consistent values for the planet-to-host mass ratio $q$. 
For example, the well-known central caustic ``close-wide'' degeneracy, which is commonly observed in the high-magnification events \citep{Griest1998,Dominik1999,An2005_CloseWideDegeneracy} \HL{in which the source approaches} the central caustic, and the ``inner-outer'' degeneracy \citep{Gaudi1997_InnerOuterDenegeracy} where the source cross or passes near the planetary caustic. These degeneracies can in fact be unified by including the source trajectory angle and the impact parameter as additional parameters \citep[e.g.,][]{Ryu2022_KMTMassProduction1, KemingZhang2021_OffsetDegeneracy}. Because such degeneracies do not \HL{have a major impact on} the measurement of $q$, they do not introduce \HL{a major} bias into sensitivity estimates.

However, 2L1S degeneracies that \HL{produce} discrepant $q$ values do exist. In Table \ref{tab:lit}, we summarize the literature-reported cases of $q$-dependent degeneracies in real microlensing events and classify them into four categories. 
The first category is the central caustic ``planet-binary'' degeneracy. Events in this group are typically high-magnification events whose light curves exhibit a double-bump structure near the peak. This feature can be explained by \HL{a} central caustic approaching in either a small-$q$ planetary model \citep{Chung2005_CentralCaustic} or a large-$q$ binary model with a stellar-mass companion.
The second category comprises events that generally show a well-separated secondary peak alongside the main peak. Such light curves can be interpreted as due to either a close or a wide planetary caustic approach or crossing \citep{Han2006_PlanetaryCaustic}.
For a given mass ratio $q$, the wide planetary caustic produces a stronger signal than the close one. Consequently, to reproduce a signal of similar magnification, the wide model often requires a smaller $q$, while the close model requires a larger $q$.
The third category, the ``central–resonant'' caustic degeneracy, includes cases that were identified early on but have only recently been recognized as a distinct class of degeneracy \citep{Ryu2022_KMTMassProduction1, Yang2022_kb210171_kb211689, ZhangJiyuan2023_kb220440}. These also occur in high-magnification events and currently lack a simple theoretical explanation. Typically, the degenerate models yield different values of both $q$ and the finite source radius $\rho$, though in some cases their $q$ values may be consistent.
The first three categories represent intrinsic degeneracies, arising from fundamental similarities in the one-dimensional magnification patterns. The fourth group includes other unclassified or coincidental degeneracies, often resulting from subtle signals and/or highly incomplete data sampling.

{
\setlength\LTcapwidth{\linewidth}
\begin{longtable*}[htb]{p{4cm}p{4cm}p{7cm}}
    \caption{Cases of degenerate multiple $q$ solutions reported in the literature}
    \label{tab:lit}
    \\ \hline \hline
    Event Name & Reference & Note \\
    \hline
    \multicolumn{3}{c}{Central caustic ``planet-binary'' degeneracy}   \\
    \hline
    OGLE-2003-BLG-340  & \citet{Jaroszynski2004_OGLEBinary_2002_2003} & equally good $+$ 2S model\\
    OGLE-2011-BLG-0526 & \citet{Choi2012_ob110526_ob110950_PlanetBinaryDegeneracy} & planetary $q$ preferred \\
    \HL{OGLE-2011-BLG-0950} & \citet{Choi2012_ob110526_ob110950_PlanetBinaryDegeneracy, Suzuki2016}; \citet{Terry2022_OB110950AO} & \HL{planetary $q$ preferred ($\Delta\chi^2<20$ from the optimized photometry), but ruled out by the Keck follow-up observations} \\
    OGLE-2012-BLG-0455 & \citet{Park2014_OB120455_PlanetBinaryDegeneracy} & planetary $q$ preferred\\
    MOA-2015-BLG-337   & \citet{Miyazaki2018_MB15337_PlanetBinaryDegeneracy} & planetary $q$ preferred\\
    KMT-2018-BLG-2164  & \citet{Gould2022_AnomalyFinder5_2018prime} & planetary $q$ preferred\\
    OGLE-2018-BLG-1554 & \citet{Gould2022_AnomalyFinder5_2018prime} & equally good\\
    KMT-2022-BLG-0371  & \citet{Han2023_PartiallyCovered} & planetary $q$ preferred\\
    KMT-2016-BLG-1243  & \citet{Shin2024_AnomalyFinder11_2016SubPrime} & binary $q$ preferred \\
    OGLE-2016-BLG-1704 & \citet{Shin2024_AnomalyFinder11_2016SubPrime} & binary $q$ preferred \\
    OGLE-2016-BLG-1408 & \citet{Shin2024_AnomalyFinder11_2016SubPrime} & planetary $q$ preferred\\
    KMT-2017-BLG-0958  & \citet{Gui2024_AnomalyFinder12_2017SubPrime} & binary $q$ preferred\\
    KMT-2023-BLG-1896  & \citet{Han2025_MB22033_KB23119_KB231896} & planetary $q$ preferred\\
    \hline
    \multicolumn{3}{c}{Planetary caustic ``close-wide'' degeneracy} \\
    \hline
    OGLE-2017-BLG-0373 & \citet{Skowron2018_OB170373} & equally good\\
    KMT-2016-BLG-0212  & \citet{Hwang2018_KB160212_PlanetaryCloseWideDegeneracy} & ``close'' preferred\\
    OGLE-2011-BLG-0173 & \citet{Poleski2018_OB110173} & equally good $+$ 2S model\\
    KMT-2017-BLG-1146  & \citet{Shin2019_KB171038_KB171146} & ``close'' preferred \\
    OGLE-2018-BLG-0596 & \citet{Jung2019_OB180596_PlanetaryCloseWideDegeneracy} & ``close'' strongly preferred \\
    KMT-2016-BLG-1107  & \citet{Hwang2019_KB161107_PlanetaryCloseWideDegeneracy} & ``close'' strongly preferred \\
    OGLE-2014-BLG-0319 & \citet{Miyazaki2022_OB140319} & ``wide'' preferred \\
    K2-2016-BLG-0005   & \citet{Specht2023_K2C9_160005} & ``wide'' preferred \\
    OGLE-2017-BLG-1806 & \citet{Zang2023_AnomalyFinder7_m4planet} &  ``close'' preferred \\
    KMT-2016-BLG-1105  & \citet{Zang2023_AnomalyFinder7_m4planet} & ``wide'' preferred \\
    KMT-2016-BLG-0625  & \citet{Shin2023_AnomalyFinder9_2016Prime} &  ``close'' preferred \\
    OGLE-2017-BLG-0448 & \citet{Zhai2024_OB170448} & Unclear \\
    \hline
    \multicolumn{3}{c}{``Central-resonant'' caustic degeneracy, or ``continuous $\rho$'' degeneracy}   \\
    \hline
    OGLE-2011-BLG-0251 & \citet{Kains_ob110251_CentralResonantDegeneracy} & ``central'' small-$\rho$ larger $q$ preferred\\
    MOA-2011-BLG-262   & \citet{Bennett2014_MB11262_GalacticModel, Terry2025_MB11262AO} & ``resonant'' small-$\rho$ larger $q$ preferred,  but  disfavored by the Keck follow-up observation\\
    OGLE-2018-BLG-0740 & \citet{Han2019_OB180740_CentralResonantDegeneracy} & ``central'' small-$\rho$ larger $q$ preferred\\
    KMT-2021-BLG-1391  & \citet{Ryu2022_KMTMassProduction1} & ``resonant'' small-$\rho$ larger $q$ preferred  \\
    KMT-2021-BLG-1253  & \citet{Ryu2022_KMTMassProduction1} & consistent $q$, ``resonant'' small-$\rho$ slightly preferred \\
    KMT-2021-BLG-0171  & \citet{Yang2022_kb210171_kb211689} & ``central'' small-$\rho$ larger $q$ preferred \\
    KMT-2021-BLG-1689  & \citet{Yang2022_kb210171_kb211689} & ``central'' high-$\rho$ larger $q$ preferred\\
    \HL{KMT-2023-BLG-0416}  & \citet{Han2024_KB230416_KB231454_KB231642} & consistent $\rho$, ``resonant'' smaller $q$ preferred \\
    \hline
    \multicolumn{3}{c}{Other degeneracy} \\
    \hline
    OGLE-2002-BLG-055  & \citet{GaudiHan2004_OB02055} & continuous $q$ degeneracy \\
    MOA-2007-BLG-192   & \citet{Bennett2008_MB07192} & all are planet solutions \\
    \HL{OGLE-2017-BLG-0173} & \citet{OB170173} & \HL{planetary wide caustic ``inner-outer-crossing'', low-$q$ ``crossing'' slightly preferred} \\
    MOA-bin-29         & \citet{Kondo2019_MOAbin29} & all are planet solutions \\
    OGLE-2015-BLG-1771 & \citet{Zhang2020_OB151771} & 3 planet solutions\\
    KMT-2019-BLG-1339  & \citet{Han2020_KB191339} & planet solution preferred\\
    KMT-2018-BLG-1025  & \citet{Han2021_KB181025} & 3 planet solutions\\
    \HL{KMT-2019-BLG-0414}  & \citet{Han2022_KB170673_KB190414} & 4 planet solutions \\
    KMT-2018-BLG-1497  & \citet{Jung2022_AnomalyFinder6} & 3 planet solutions $+$ 2S model\\
    KMT-2018-BLG-1714  & \citet{Jung2022_AnomalyFinder6} & 3 planet solutions  $+$ 2S model\\
    KMT-2016-BLG-1751  & \citet{Shin2023_AnomalyFinder9_2016Prime} & larger-$q$ preferred\\
    KMT-2016-BLG-1855  & \citet{Shin2023_AnomalyFinder9_2016Prime} & 6 planet solutions $+$ 2S model\\
    OGLE-2017-BLG-1694 & \citet{Ryu2024_AnomalyFinder10_2017Prime} & low-$q$ preferred $+$ 2S model\\
    \HL{KMT-2023-BLG-1454}  & \citet{Han2024_KB230416_KB231454_KB231642} & 4 planet solutions \\
    MOA-2019-BLG-421   & \citet{Yang2024_pysis5_RAMP1} & Unclear  \\
    MOA-2016-BLG-526   & \citet{Shin2024_AnomalyFinder11_2016SubPrime} & larger-$q$ preferred\\
    KMT-2021-BLG-2609  & \citet{Han2024_KB212609_KB220303} & small-$q$ preferred\\
    \HL{OGLE-2018-BLG-0421} & \citet{Yang2025_RAMP2} & 6 binary solutions + 9 planetary solutions, planetary-$q$ slightly preferred \\
    OGLE-2015-BLG-1609 & \citet{Mroz2025_OB151609} & Unresolved\\
    \hline
    \multicolumn{3}{c}{NOTE. ``2S'' denotes the binary-source models.}
\end{longtable*}
}

In actual planet detections, signals that exhibit strong degenerate models with discrepant $q$ measurements \HL{may not be} regarded as confirmed planetary signals and are therefore \HL{specially treated or} excluded from statistical samples. 
However, sensitivity calculations typically do not account for this effect due to computational complexity. 
\HL{For example, OGLE-2011-BLG-0950 \citep{Choi2012_ob110526_ob110950_PlanetBinaryDegeneracy, Suzuki2016} is an ambiguous event in the \citet{Suzuki2016} sample for which both $q\sim0.4$ stellar-binary models and $q\sim5\times10^{-4}$ planetary models can explain the light curve. The models are then weighted with priors from the $\Delta\chi^2$ and the Galactic model and then added to the statistical sample. However, such treatment is not included in the sensitivity calculation.
For the \citet{Zang2025Science_KMT_MassRatioFunction} sample, the events having $\Delta\chi^2<10$ models with discrepant mass-ratio measurements of $\Delta\log q>0.25$ are considered ambiguous events and are directly excluded. The process is also ignored in the sensitivity calculation.}
The omission may lead to an overestimation of the sensitivity and, consequently, an underestimation of the planet occurrence rate. Moreover, the resulting bias could vary with the mass ratio $q$, potentially distorting the inferred shape of the mass-ratio function. 

Beyond the 2L1S interpretations, single-lens binary-source (1L2S) models could also explain some candidate planetary signals \citep{Gaudi1997_InnerOuterDenegeracy, SangtarashYee2025_1L2S}. However, 1L2S models are computationally more straightforward to sample and have already been incorporated into standard sensitivity calculations \citep[e.g.,][]{Zang2025Science_KMT_MassRatioFunction}. We therefore focus on the less-explored effects of 2L1S degeneracy in this study.

In this work, we examine the influence of 2L1S degeneracies on planetary sensitivity calculations by simulating five groups of typical events. The paper is organized as follows: Section~\ref{sec:method} outlines the simulation methodology, including the configuration of typical event parameters, light curve generation, and the global search for possible models and the model comparison. In Section~\ref{sec:result}, we present several examples of degeneracies identified through the simulation and \HL{evaluate} their impact on sensitivity. Finally, Section~\ref{sec:dis} discusses the implications for the statistical analysis of the microlensing planet mass-ratio function and the potential applications.

\section{Simulation Methodology and Setup}\label{sec:method}

\subsection{Preamble}
\label{sec:method_pre}

Accurately estimating planetary sensitivity requires closely replicating the procedures used in actual planet detections, including the identification of anomalous signals, global parameter search, and model comparison.
We follow this procedure in our study. Specifically, we define the sensitivity of an event as
\begin{equation}
    \e(s,q,\alpha) = \eanom(s,q,\alpha) \cdot \edeg(s,q,\alpha),
\end{equation}
where $s$ denotes the secondary-to-primary lens separation in units of the Einstein radius, $q$ the mass ratio, and $\alpha$ the angle between the source-lens relative trajectory and the primary-to-secondary lens axis. Here, $\eanom$ and $\edeg$ represent the \HL{boolean} probabilities that a planetary signal with given parameters $(s, q, \alpha)$ is detected and correctly characterized \HL{($q$ uniquely determined)}, respectively. The overall sensitivity $\e$ is thus the product of $\eanom$ and $\edeg$.

\begin{figure}
    \centering
    \includegraphics[width=0.35\columnwidth]{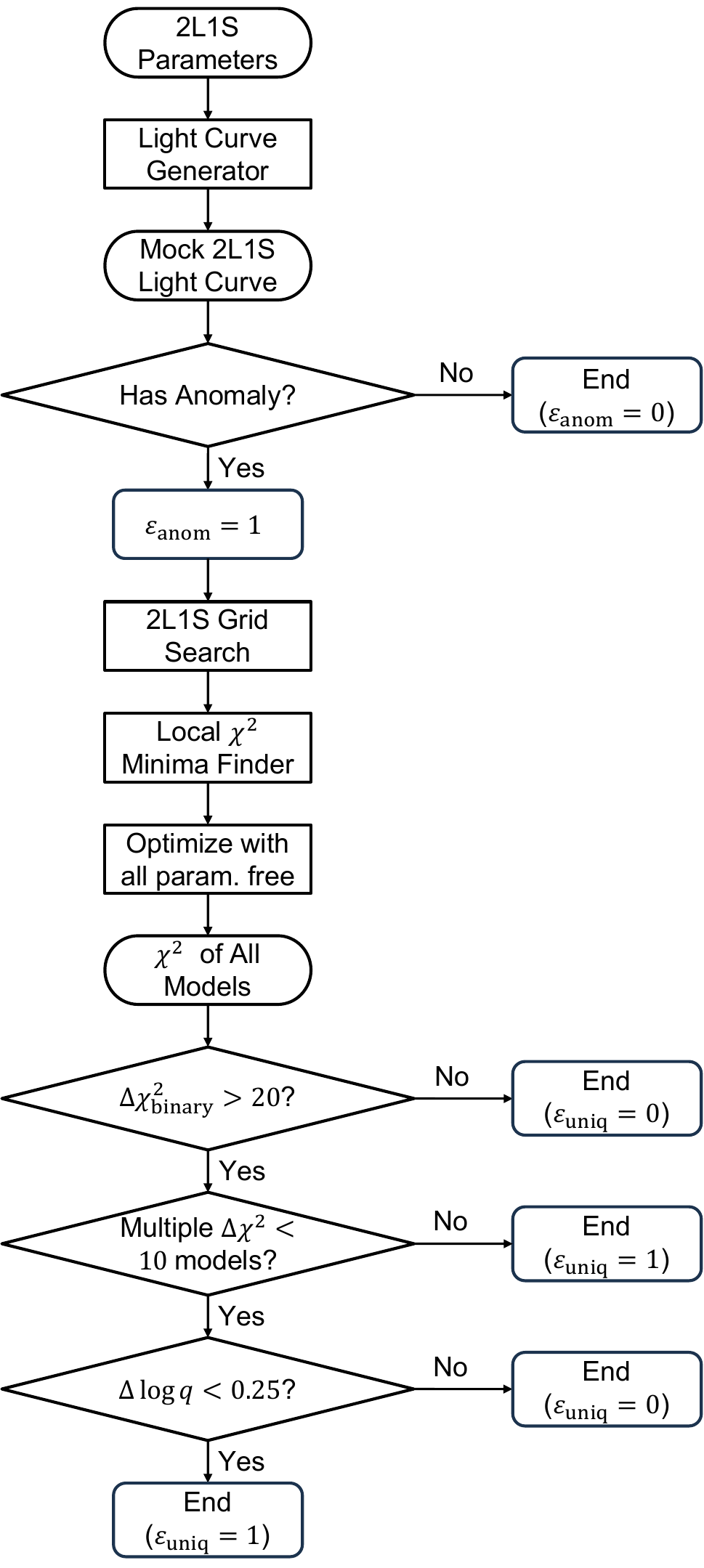}
    \caption{Flow chart of the sensitivity calculation procedure.}
    \label{fig:flowchart}
\end{figure}

Figure~\ref{fig:flowchart} illustrates the overall workflow of our sensitivity calculation. The following sections describe these steps in detail.

\subsection{Light Curve Generation}
\label{sec:genlc}

The magnification as a function of time, $A(t)$, for a 2L1S microlensing event is characterized by at least seven parameters. In addition to $(s, q, \alpha)$ as introduced above, the remaining four parameters are: $t_0$, the time at which the source is closest to the magnification center of the lens system; $u_0$, the impact parameter in units of the Einstein radius; $\tE$, the Einstein radius crossing time (also known as the microlensing timescale); and $\rho$, the angular radius of the source star in units of the Einstein radius.  
Additionally, to convert $A(t)$ into a measurable light curve, two magnitude parameters, $I_{\rm S}$ and $I_{\rm B}$, are required to represent the source star magnitude and the blended light, respectively.

To investigate degeneracy effects and cover typical events, we sample five groups of microlensing events, they are giant-source low-magnification (GSLM), giant-source high-magnification (GSHM), dwarf-source low-magnification (DSLM), dwarf-source high-magnification (DSHM), and dwarf-source extreme-magnification (DSEM) events. 
Giant sources have larger source radii $\rho$, while dwarf sources have smaller $\rho$. The impact parameter $u_0$ varies from larger values for low-magnification events to smaller values in high- and extreme-magnification events.
Detailed parameter settings for each group are provided in Table \ref{tab:setup}.
These settings allow us to systematically explore how degeneracy properties depend on different types of events. Moreover, different statistical samples are expected to be influenced to varying degrees.
For example, early statistical samples of microlensing planets were constructed primarily from follow-up observations of high-magnification and extreme-magnification events \citep{Gould2010_mufun_stat,Cassan2012Nature}. 
\HL{Subsequent studies adopted a hybrid strategy incorporating both survey and follow-up data \citep[for example,][]{Suzuki2016}. These samples tend to be dominated by high-magnification events but also include some low-magnification events.
More recent samples based on survey-only data \citep{ShvartzvaldYossi2016_Wise, Poleski2021_WideOrbitStatistic, Zang2025Science_KMT_MassRatioFunction} contain a broader range of event types.}
\HL{Given that the observational strategies of follow-up and hybrid modes often involve complex, irregular cadences, we focus here on the pure-survey mode in our simulation for simplicity. Nevertheless, the results can still provide some understanding for other strategies.}

\begin{table}
    \centering
    \caption{Simulation Setup Parameters}
    \renewcommand{\arraystretch}{1.2}
    \label{tab:setup}
    \begin{tabular}{cccc}
        \hline
        Group & $I_{\rm S}$ & $\log\rho$ & $u_0$ \\ \hline
        GSLM &    $17.0$   &   $-2.0$   & $0.50$ \\
        GSHM &    $17.0$   &   $-2.0$   & $0.02$ \\
        DSLM &    $21.0$   &   $-2.8$   & $0.20$ \\
        DSHM &    $21.0$   &   $-2.8$   & $0.02$ \\
        DSEM &    $21.0$   &   $-2.8$   & $0.002$ \\
        \hline
    \end{tabular}
    \begin{tablenotes}
        \centering
        \item{NOTE. D/GS = Dwarf/Giant Source; L/H/EM = Low/High/Extreme Magnification.
        Other parameters, $t_0=0$, $\tE=15~{\rm d}$, and $I_{\rm B}=19.5$ are the same for all groups.}
    \end{tablenotes}
\end{table}

We then sample a series of values of $(s, q, \alpha)$ within each group to investigate degeneracy properties as functions of these parameters. 
To reduce the number of sampling points and computational cost, we sample $s$ and $q$ in the ``$(k,h)$'' parameter space proposed by \citet{khgrid_Hall2024}. This parameterization allows more efficient sampling of magnification configurations.
The parameter $k$ reflects the width of the central caustic, while $h$ is nearly orthogonal to $k$ in the $(\log s, \log q)$ space.
Specifically, they are defined as \citep{khgrid_Hall2024}:
\begin{equation}
    k(s,q)=
    \begin{cases}
        \log q + 2.3& |\log s|\leq\log s_{\rm ref}\\
        \log\left[\frac{4q}{(s-s^{-1})^2}\right]& |\log s|>\log s_{\rm ref}
    \end{cases}
    \label{eq:kdef}
\end{equation}
where $\log s_{\rm ref}=0.03$ is an artificial boundary to avoid divergence, and
\begin{equation}
    h(s,q)=
    \begin{cases}
        35 \log s& |\log s|\leq\log s_{\rm ref},\\
        \frac{\log s}{|\log s|}\left[\sqrt{(\log s - \log s_{\rm ref})^2+(\log q-\log q_{\rm ref})^2}+C\right]& |\log s|>\log s_{\rm ref},
    \end{cases}
    \label{eq:hdef}
\end{equation}
with $\log{q_{\rm ref}} = k(s,q)-k(s_{\rm ref},1)$ and $C = 35\log s_{\rm ref}$. 
We sample 16 equally spaced values of $k$ in the range $-3.7 \leq k \leq 2.3$, and 15 equally spaced values of $h$ in each of the intervals $-1.05 \leq h \leq 1.05$ (where $C \cdot \log{s_{\rm ref}} = 1.05$), $-6 \leq h < -1.05$, and $1.05 < h \leq 6$. 
After filtering out \HL{grid points} with $|\log s| > 1.5$ or $\log q > 0$, a total of 533 grid points remain. 
Note that we generate light curves only for grid points with $\log q < -1.5$ to focus on planetary sensitivity, resulting in 368 $(k, h)$ (or equivalently $(s, q)$) points. However, the parameter search (Section \ref{sec:grid}) is performed over all 533 grid points to account for potential stellar-mass binary solutions with large $\log q$.

For $\alpha$, we sample 60 equally spaced values over $0^\circ \leq \alpha < 360^\circ$ for high- and extreme-magnification events, but increase this to 180 values for low-magnification events. This distinction is made because planetary signals in high-magnification events primarily arise from source approaches to the central caustic and are less sensitive to $\alpha$, whereas those in low-magnification events typically result from planetary caustic crossings that \HL{are} more sensitive to $\alpha$ and \HL{require} denser sampling. 

We compute the 2L1S magnification using the \texttt{VBBinaryLensing} package \citep{Bozza2010_VBBL,Bozza2018_VBBL}. 
The light curve spans from $-45\,{\rm d}$ to $+45\,{\rm d}$ ($-3\tE$ to $+3\tE$, \HL{where $\tE=15~{\rm d}$}), and the cadence \HL{is} set to be $\Gamma=4\,{\rm hr}^{-1}$ to simulate the \HL{pure-survey} observational strategy of the Korean Microlensing Telescope Network \citep[KMTNet,][]{Kim2016_KMT}. \HL{The sample points are ideal, continuous over the time interval, and are free of weather conditions and moon phase. We note that this may underestimate the effect of degeneracy because some degeneracy cases in Table~\ref{tab:lit} can be resolved if the time sampling were continuous.}

\begin{figure}
    \centering
    \includegraphics[width=0.5\columnwidth]{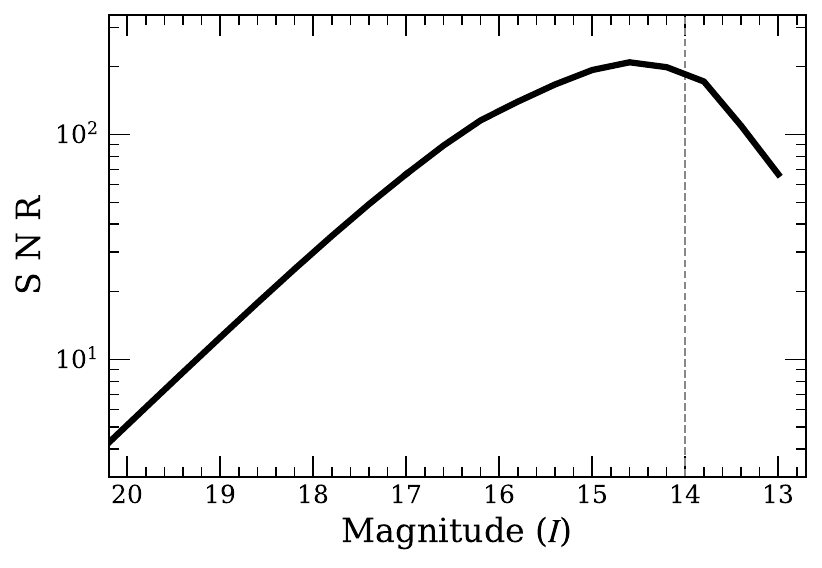}
    \caption{The SNR curve used in the light curve generation, obtained from applying the \citet{Qian2025_KMTFFP1} pipeline to the KMTNet-SAAO 2018 observations.}
    \label{fig:snr}
\end{figure}

For the noise model, we adopt the median SNR curve derived from KMTNet observations during the 2018 season using the pipeline described in \citet{Qian2025_KMTFFP1}. 
Specifically, we use the SNR curve from the telescope at the South African Astronomical Observatory (SAAO) as representative of the median KMTNet performance. 
The SNR curve is shown in Figure \ref{fig:snr}.
Note that we do not add noise directly to the light curve, instead, we use the SNR curve to assign an error bar to each data point, which is then used in subsequent analysis.
Statistically, in terms of $\chi^2$ evaluation, both approaches are equivalent, but the latter offers better numerical stability.

In summary, we generate 22,080 planetary 2L1S light curves for each of the three high- and extreme-magnification groups, and 66,240 for each of the two low-magnification groups.

\subsection{Anomaly Identification}
\label{sec:eaf}

After generating the light curves, we first search for the anomalous signals. 
An anomaly is defined as a statistically significant deviation from the single-lens single-source (1L1S) model.
To identify such signals, we fit the generated 2L1S light curves with 1L1S models and examine the residuals. 
We adopt the \href{https://github.com/hongjingy/easyAnomalyFinder}{\texttt{easyAnomalyFinder}} package to search for residual signals \citep{Yang2025_RAMP2}. This package scans the light curve residuals using a series of time windows and applies multiple statistical tests to determine whether a signal is sufficiently significant within each window.

In the \texttt{easyAnomalyFinder} configuration, we set the overall significance threshold to be $4 \sigma$ and require a minimum $\Delta\chi^2$ to the 1L1S model to be $\Delta\chi^2_{\rm 1L1S}>\Delta\chi^2_{\rm min}=80$. \HL{Because} the simulated light curves are noise-free, we disable most data quality checks. However, we enable the $\chi^2$ dominance check and the smoothness check to exclude very short or poorly sampled signals. In addition, we require at least one data point to deviate by $1\sigma$ to filter out \HL{anomalies that are too subtle}.

For each light curve (or its residual sequence), \texttt{easyAnomalyFinder} returns a judgment indicating whether any detectable anomalies exist. This corresponds to a boolean value of $\eanom$:
\begin{equation}
    \eanom(s,q,\alpha)=\left\{
    \begin{aligned}
        1, & & {\rm has\ detectable\ anomaly};  \\
        0, & & {\rm no\ detectable\ anomaly}.
    \end{aligned}
    \right.
\end{equation}

After applying the anomaly search, the number of light curves with detectable anomalies for each group (GSLM, GSHM, DSLM, DSHM, DSEM) is (11432, 10907, 10481, 11732, 13529), respectively. 
Because light curves with $\eanom = 0$ exhibit no sensitivity to planets, only those with $\eanom = 1$ are processed further in the subsequent steps.

\subsection{Global 2L1S Model Search}
\label{sec:grid}
As noted in Section~\ref{sec:intro}, a global search across the 2L1S parameter space is essential to identify all possible models.
We implement this through a grid search, which is a standard method for characterizing microlensing signals.
Unlike most previous studies that use linear grids in the $(\log q, \log s)$ space, our grid adopts the $(k, h)$ parameterization introduced in Section \ref{sec:genlc}.
Furthermore, we extend the grid to include regions with $\log q > -1.5$ (for a total of 533 grid points), which were omitted in Section~\ref{sec:genlc}, in order to explore potential degeneracies between planetary signals and stellar binary solutions.

For each $(s, q)$ grid point, we fix these two parameters along with $\rho$, and use Markov Chain Monte Carlo (MCMC) sampling via \texttt{emcee} \citep{emcee} to explore the $\chi^2$ surface of the remaining parameters.
We initialize eight uniformly distributed values of $\alpha$ (from $0^\circ$ to $360^\circ$) per grid point, while other parameters are set to their true input values.
To ensure thorough exploration of the parameter space, we employ a gradually cooling ``hot chain'' strategy, \HL{that is}, the $\chi^2$ is scaled by a factor that gradually decreases from $50/\Delta\chi^2_{\rm 1L1S}$ to $1$ over 1000 steps, where $\Delta\chi^2_{\rm 1L1S}$ is obtained from Section~\ref{sec:eaf}.
This is followed by 500 steps of Nelder-Mead optimization to refine the best model at each grid point.

During the grid search, we compute \HL{the} 2L1S magnification by interpolating pre-computed magnification maps (Zhang et al., in prep).
This reduces \HL{the} computational costs of the grid search per light curve from $\sim 400$ to $\sim 4\,{\rm CPU\cdot hr}$, making such extensive grid searches that were previously limited by computational resources now feasible.

We successfully performed the grid search for all 58,081 light curves with $\eanom = 1$, obtaining $\chi^2$ samplings for all 2L1S parameters. The next step is to identify local $\chi^2$ minima as candidate solutions and further optimize them.

\subsection{Local Minima Locating and Optimizing}
\label{sec:local_minima}
Conventionally, in microlensing planetary signal \HL{analyses}, local $\chi^2$ minima are identified visually by projecting grid search results onto the $(\log s, \log q, \alpha)$ planes.
However, this interactive, human-dependent approach becomes impractical for the large event samples considered in this study.
We therefore develop an automated algorithm for detecting local minima.

The algorithm searches for local minima in $(k, h)$ space, as the magnification patterns are more uniformly sampled under this parameterization.
First, we project the $(k, h, \alpha)$ results onto the $(k, h)$ plane by taking, for each $(k, h)$ grid point, the minimum $\chi^2$ across all sampled $\alpha$ values.
Second, for every $(k, h)$ grid point, we identify all neighboring points within a Euclidean distance $R \leq 0.5$ in the $(k, h)$ parameter space.
Then a grid point is classified as a local minimum in $(k, h)$ if its $\chi^2$ is lower than that of all its neighbors.

Lastly, for each such $(k, h)$ local minimum, we group all associated $\alpha$ values that lie within $5^\circ$ of each another and retain only the $\alpha$ value with the smallest $\chi^2$ in each group. 
This procedure identifies all local $\chi^2$ minima across the $(k, h, \alpha)$ space, or equivalently, the $(s, q, \alpha)$ space, for each simulated event.
In total, 2,705,397 local $\chi^2$ minima (candidate solutions) are identified.

Each candidate solution then undergoes a final refinement via 500 steps of Nelder–Mead optimization, with all parameters (including $s$, $q$, and $\rho$) unconstrained.

\subsection{Model Comparison}
\label{sec:model_compare}
After completing the previous steps, we obtain one or more candidate models for each simulated planetary light curve.
We then compare these models to assess whether the planetary signal can be reliably characterized, as illustrated in Figure \ref{fig:flowchart}.

First, we examine whether any of the remaining models are consistent with a stellar-mass binary companion. 
If a light curve has any solution with $\log q\geq-1.5$ and $\Delta\chi^2_{\rm binary}=\chi^2_{{\rm min},\,\log q>-1.5}\leq20$, we classify it as a planet–binary degenerate case and assign its $\edeg = 0$.
Note that, \HL{because} the generated light curves are noise-free, the true model has $\chi^2 = 0$, and for any other model, $\Delta\chi^2$ equals its $\chi^2$ value.
Light curves not satisfying this condition proceed to the next examination.

Next, we evaluate whether the light curve can be explained by multiple planetary models with significantly different mass ratios.
Specifically, we identify all models with $\Delta\chi^2 < 10$ and check whether any pair exhibits a discrepancy $\Delta\log q \geq 0.25$ \citep[e.g.,][]{Zang2025Science_KMT_MassRatioFunction}.
If such a pair exists, we conclude that the light curve suffers from multi-planet degeneracy and cannot be uniquely characterized, and accordingly, $\edeg = 0$. 
If only one planetary model remains, or if all such models agree within $\Delta\log q < 0.25$, we consider the planetary signal to be solidly characterized and assign $\edeg = 1$.

After applying this model comparison to all simulated events, we find 1,052 cases with $\edeg = 0$ among all the 58,081 light curves where $\eanom = 1$.

\subsection{Simulation Summary}
\label{sec:simu_summary}

We successfully simulated 198,720 light curves and performed a systematic analysis including anomaly identification, global model search, local optimization, and model comparison. 
Table \ref{tab:simu_summary} summarizes the overall procedure and the associated computational costs. 
Through this process, we derive the sensitivity both with ($\e = \eanom \cdot \edeg$) and without ($\eanom$) accounting for 2L1S degeneracies, as a function of $(k, h, \alpha)$ or equivalently $(\log s, \log q, \alpha)$. 
In the next section, we present example light curves and derive the impact of these degeneracies on planetary sensitivity.

\begin{table}[]
    \centering
    \caption{Simulation Summary}
    \renewcommand{\arraystretch}{1.2}
    \label{tab:simu_summary}
    \begin{tabular}{|clllll|l|}
        \hline
        \multicolumn{1}{|l|}{}                       & \multicolumn{1}{l|}{DSEM}   & \multicolumn{1}{l|}{DSHM}   & \multicolumn{1}{l|}{DSLM}   & \multicolumn{1}{l|}{GSHM}   & GSLM                        & Computation Cost (CPU$\cdot$hr) \\ \hline
        \multicolumn{1}{|c|}{Light Curve Generation} & \multicolumn{1}{c|}{22,080} & \multicolumn{1}{c|}{22,080} & \multicolumn{1}{c|}{66,240} & \multicolumn{1}{c|}{22,080} & \multicolumn{1}{c|}{66,240} &                            \\ \hline
        \multicolumn{6}{|c|}{Anomaly Identification}                                                                                                                                                               &    276                        \\ \hline
        \multicolumn{1}{|c|}{Anomaly ($\eanom=1$)}                & \multicolumn{1}{c|}{13,529} & \multicolumn{1}{c|}{11,732} & \multicolumn{1}{c|}{10,481} & \multicolumn{1}{c|}{10,907} & \multicolumn{1}{c|}{11,432} &                            \\ \hline
        \multicolumn{6}{|c|}{Global 2L1S Grid Search}                                                                                                                                                           &    $2.59\times 10^{5}$                        \\ \hline
        \multicolumn{1}{|c|}{Local Minima}           & \multicolumn{1}{l|}{697,031}       & \multicolumn{1}{l|}{590,160}       & \multicolumn{1}{l|}{513,362}       & \multicolumn{1}{l|}{361,154}       &            543,690                 &                            \\ \hline
        \multicolumn{6}{|c|}{Nelder-Mead Optimization with all Parameters Free}                                                                                                                                                                     &         $7.51\times10^{4}$                   \\ \hline
        \multicolumn{6}{|c|}{Total}                                                                                                                                                  &    $3.34\times 10^{5}$                        \\ \hline
    \end{tabular}
\end{table}

\section{Results}\label{sec:result}

\subsection{Example Degenerate Events}
\label{sec:result_case}
In this section, we present several examples of the degeneracy cases identified in our simulation and discuss their properties.

\subsubsection{The ``Planet-binary'' Degeneracy}
\label{sec:example:planet-binary}

\begin{figure*}
    \centering
    \includegraphics[width=0.75\columnwidth]{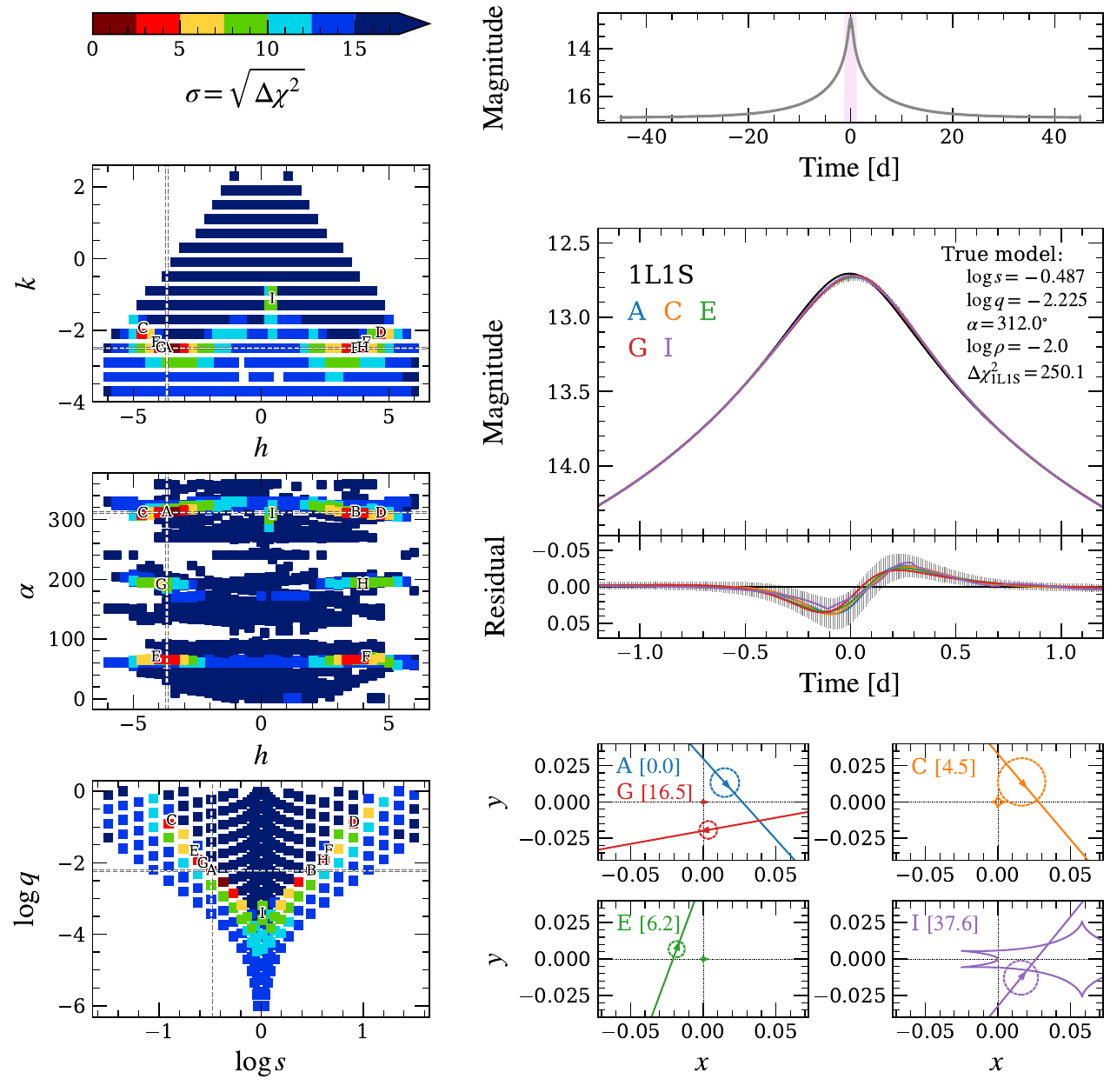}
    \caption{An example of a simulated $\edeg=0$ event from the GSHM group, affected by the ``planet-binary'' degeneracy.
    Left panels show the grid search results projected onto \HL{the} $(h,k)$, $(h,\alpha)$, and $(\log s, \log q)$ planes, with identified local minima labeled. 
    The top right panel displays the full simulated light curve, and the middle right panels show a zoomed-in view of the anomalous region along with several selected optimized models and residuals relative to the 1L1S model. 
    The bottom right panels present the source-lens trajectories and the caustic geometries for each selected model. Dashed circles indicate the source sizes, and the $\Delta\chi^2=\chi^2-\chi^2_{\rm best}$ value for each model is labeled in ``[~]'' brackets.
    In this case, Models (A, B), (C, D), (E, F), and (G, H) form ``close-wide'' degenerate pairs; thus, only the ``close'' models are shown for clarity. The central caustics of Models A and G are visually indistinguishable and are therefore shown together in one panel.
    Although models G, H, and I are excluded in this example due to their large $\Delta\chi^2$, they could survive under the conditions of lower effective cadence or lower SNR.}
    \label{fig:lc_exp1}
\end{figure*}

The first example is a ``planet-binary'' degeneracy case from the GSHM group. The light curve, grid search results, identified models, and their caustic-trajectory geometries are shown in Figure \ref{fig:lc_exp1}.

The light curve shows a subtle asymmetric feature around the peak region. 
Thirteen local minima are identified in the grid search. After optimization, they converged to nine distinct solutions, labeled A through I. Solutions (A, B), (C, D), (E, F), and (G, H) each form a ``close-wide'' pair. 
The best model, A, corresponds to the true input model with a planetary mass ratio $\log q \sim -2.2$, while model \HL{C} (and its wide counterpart D) \HL{have} a stellar binary mass ratio of $\log q \sim -0.83$. 
The $\chi^2$ values of these models differ by only 4.5, indicating that the event would not be classified as a secure planetary detection.
We find that running a long Markov Chain Monte Carlo (MCMC) with all 2L1S parameters free allows models C and D to converge toward A and B, respectively. However, this does not change the conclusion; even if C and D are not fully distinct local minima, the combined fact of large $\log q$ and low $\Delta\chi^2$ still challenges the confidence in a uniquely planetary interpretation.

In practice, planetary and binary models can sometimes yield significantly different measurements of the source size $\rho$. For example, in the case of OGLE-2011-BLG-0950 \citep{Choi2012_ob110526_ob110950_PlanetBinaryDegeneracy, Suzuki2016}, the planetary models give $\rho \approx 4.2 \times 10^{-3}$, while the binary models yield $\rho \approx 0.4 \times 10^{-3}$ and $\rho \approx 0.8 \times 10^{-3}$ for the close and wide configurations, respectively. Such an order-of-magnitude discrepancy in $\rho$ makes it possible to break the degeneracy. Indeed, follow-up high-resolution imaging by \citet{Terry2022_OB110950AO} allowed the planetary models to be ruled out for this event.
In contrast, for the simulated example shown in Figure~\ref{fig:lc_exp1}, the measured $\rho$ values do not differ substantially: The planetary models (A and B) have $\rho = 0.01$ (the true input value), and the binary models have $\rho \approx 0.015$. Such a small difference would be challenging to distinguish them even with high-resolution follow-up observations.

Unexpectedly, we also identify an additional candidate solution, ``I'', for this light curve. 
It corresponds to a very sharp local minimum in $(\log s, \log q)$ space that could easily be missed in evenly spaced $(\log s, \log q)$ grid searches. 
In this example, Solution I can be robustly excluded due to its high $\Delta\chi^2 = 37.6$. However, it might survive under a poorer light curve coverage \HL{or} lower SNR conditions.

\subsubsection{The Planetary Caustic ``Close-wide'' Degeneracy}
\label{sec:example:planet-close-wide}

\begin{figure*}
    \centering
    \includegraphics[width=0.75\columnwidth]{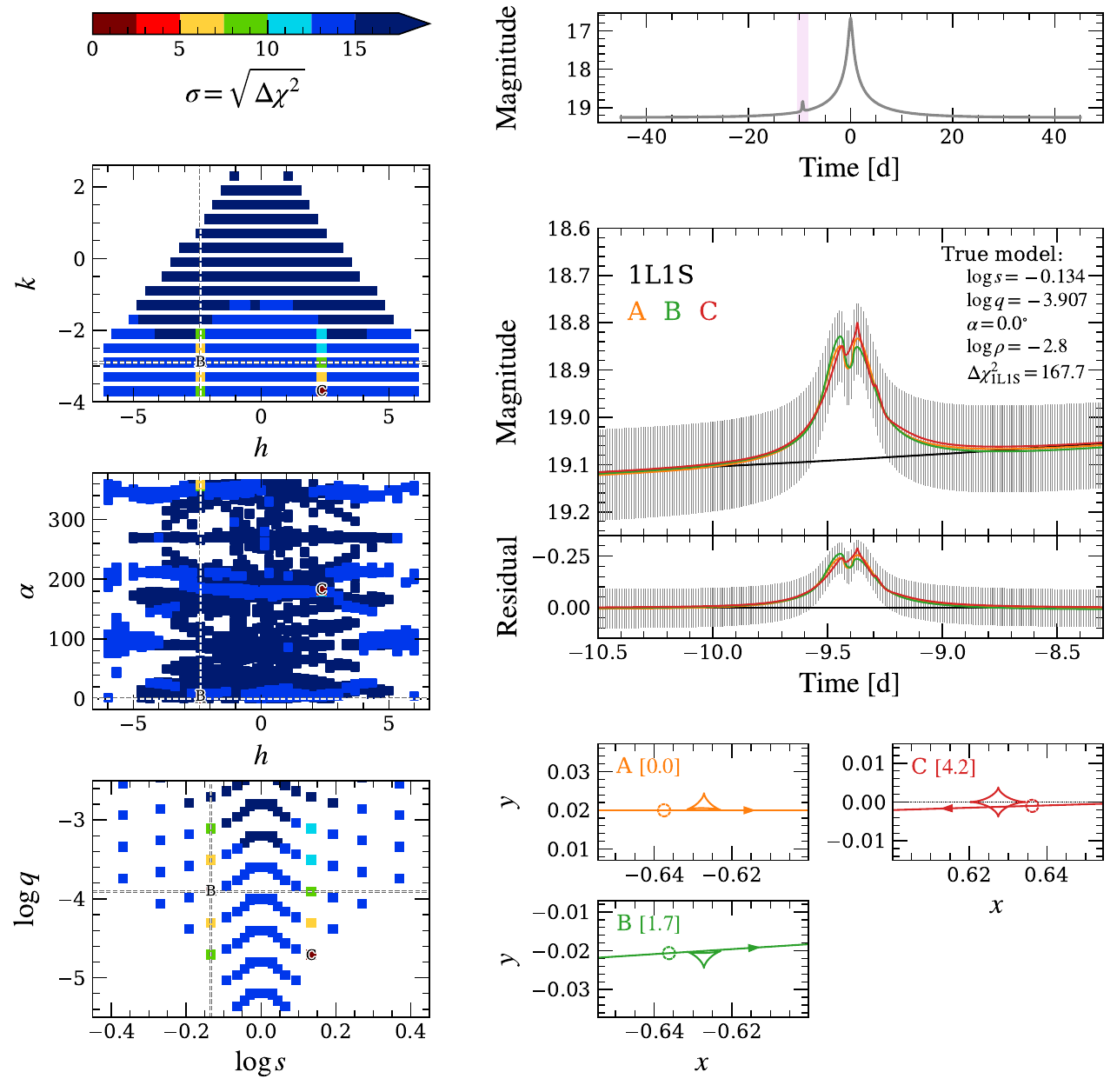}
    \caption{Same as Figure \ref{fig:lc_exp1}, but for an example $\edeg=0$ event from the DSHM group that \HL{is} affected by the planetary caustic ``close-wide'' degeneracy.
    Models A and B are visually indistinguishable on the grid search plots but represent different geometries.
    }
    \label{fig:lc_exp2}
\end{figure*}

The second example is a case of planetary caustic ``close-wide'' degeneracy from the DSHM group. 
The light curve, grid search results, identified models, and their caustic-trajectory geometries are shown in Figure \ref{fig:lc_exp2}. 

The light curve shows a shorter, isolated peak in addition to the main peak, indicating a planetary caustic crossing. This feature can be produced when the source crosses one of the two ``close'' planetary caustics for $s<1$ (Models A and B), or the single ``wide'' planetary caustic for $s>1$ (Model C) \citep{Han2006_PlanetaryCaustic}. The light curve and configuration are similar to some real events, \HL{for example}, OGLE-2018-BLG-0383 \citep{Wang2022_AnomalyFinder3_kb180900_ob180383}. 
To produce similar signals, the ``close'' configuration requires a larger $q$, while the ``wide'' configuration corresponds to a smaller $q$. It is confirmed in this example \HL{for which} the ``close'' models A and B have $\log q\approx-3.9$ and the ``wide'' model C has $\log q\approx-4.7$. The mass ratios measured by these two configurations differ by nearly an order of magnitude. All three models yield similarly good fits, with $\Delta\chi^2 < 5$, and thus \HL{they} cannot be ruled out. This event is therefore assigned $\edeg = 0$ due to the presence of multiple solutions with inconsistent mass ratio estimates.

Note that Models A and B correspond to overlapping local minima in \HL{the} $(k, h)$ or $(\log s, \log q)$ plane and differ only by about $3.7^{\circ}$ in $\alpha$. Consequently, only one local minimum (A) was identified by our search algorithm (Section \ref{sec:local_minima}), which groups solutions within $\Delta\alpha < 5^\circ$.
Nevertheless, this does not affect the conclusion that the event exhibits significant degeneracy and thus has $\edeg = 0$.


\subsubsection{The ``Central-resonant'' Caustic Degeneracy}
\label{sec:example:central-resonant}
\begin{figure*}
    \centering
    \includegraphics[width=0.75\columnwidth]{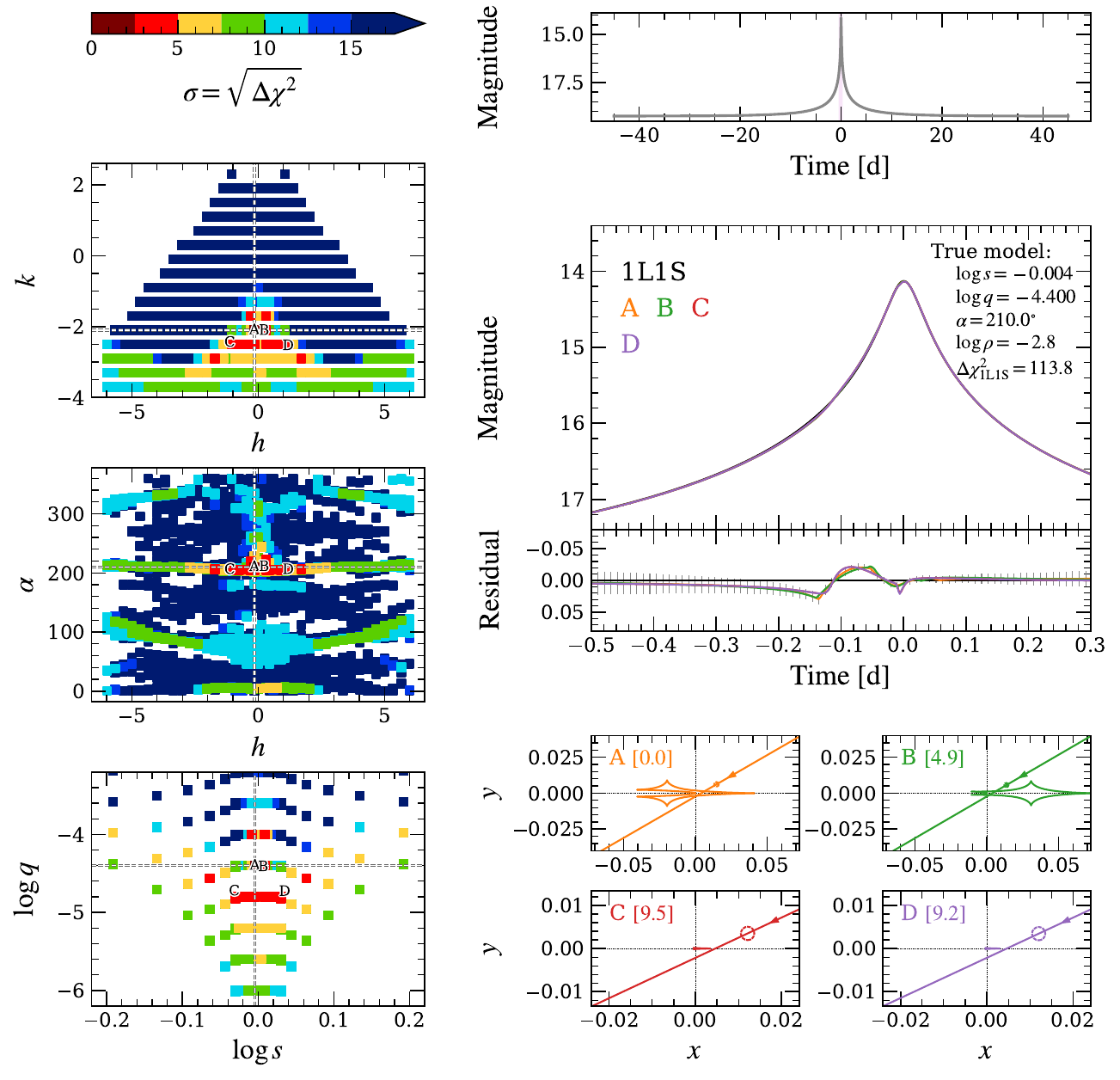}
    \caption{Same as Figure \ref{fig:lc_exp1}, but for an example $\edeg=0$ event from the DSEM group that affected by the ``central-resonant'' degeneracy.}
    \label{fig:lc_exp3}
\end{figure*}

The third example is a ``central-resonant'' caustic degeneracy case from the DSEM group. The corresponding figures are shown in Figure \ref{fig:lc_exp3}. The light curve exhibits a subtle bump relative to the 1L1S model near the peak. Such a feature can be explained either by a central caustic crossing/approach or a resonant caustic crossing, as identified in \citet{Ryu2022_KMTMassProduction1} and \citet{Yang2022_kb210171_kb211689}. 
In this example, Models A and B correspond to the resonant caustic configuration, while Models C and D correspond to the central caustic configuration. 
Additionally, (A, B) and (C, D) each form a ``close-wide'' degenerate pair. 
We note that the separation relations of these pairs do not strictly obey $s\leftrightarrow s^{-1}$. Specifically, $s_{A}\cdot s_{B}=1.0056$ and $s_{C}\cdot s_{D}=1.0080$. This slight deviation can be explained by the asymmetry induced by the source–lens trajectory's intersection with the binary-lens axis, which incorporates a dependence on $u_0/\cos\alpha$ \citet{Ryu2022_KMTMassProduction1,KemingZhang2021_OffsetDegeneracy,Zhang2022_OffsetDegeneracy}.

Among the known ``central–resonant'' caustic degeneracy cases listed in Table \ref{tab:lit}, the mass ratios measured from different configurations are sometimes consistent and sometimes discrepant. This behavior is also reflected in our simulations. 
In this particular example, the values are discrepant. The ``resonant'' models A and B yield $\log q=-4.40$, whereas and the ``central'' models C and D yield $\log q=-4.73$. Due to the inconsistent $q$ measurements and the small $\Delta\chi^2$ values (all below 10), we cannot confidently exclude either scenario or obtain a unique mass ratio estimate. Therefore, we classify this event as degenerate and assign $\edeg = 0$. 
Among these solutions, Model B was missed by the automatic local minima identification algorithm due to its proximity to Model A. However, \HL{as for} the previous example,  this does not affect the conclusion that the light curve lacks planetary sensitivity because of the $q$-discrepant degeneracy.

\citet{Yang2022_kb210171_kb211689} suggest that resonant solutions generally occupy a relatively small region in $(\log s, \log q, \alpha)$ parameter phase space and are less physically probable. 
\HL{Although the phase space argument typically contributes an equivalent $\Delta\chi^2\lesssim10$, it is considered stronger because it is less sensitive to the systematic (non-Gaussian) noise. Nevertheless, systematics can still influence this argument, particularly in complex events, for example, \citet{Yang2024_pysis5_RAMP1}. A statistically consistent method to combine phase-space priors with $\Delta\chi^2$ comparisons remains an open question.
In some cases, resonant models are preferred from the $\chi^2$ values, as in KMT-2021-BLG-1391, for which the resonant solution is favored by $\Delta\chi^2=5.1$ \citet{Ryu2022_KMTMassProduction1}.}

\HL{We note that there are also some cases for which the central and resonant solutions yield \HL{substantially} different values of $\rho$, such as MOA-2011-BLG-262 \citep{Bennett2014_MB11262_GalacticModel}, allowing model comparison to be reassessed through high-resolution follow-up observations \citep{Terry2025_MB11262AO}.}

\clearpage

\subsection{Degeneracy Impacts on Sensitivity}
\label{sec:result_sens}

\begin{figure*}
    \centering
    \includegraphics[width=1.0\columnwidth]{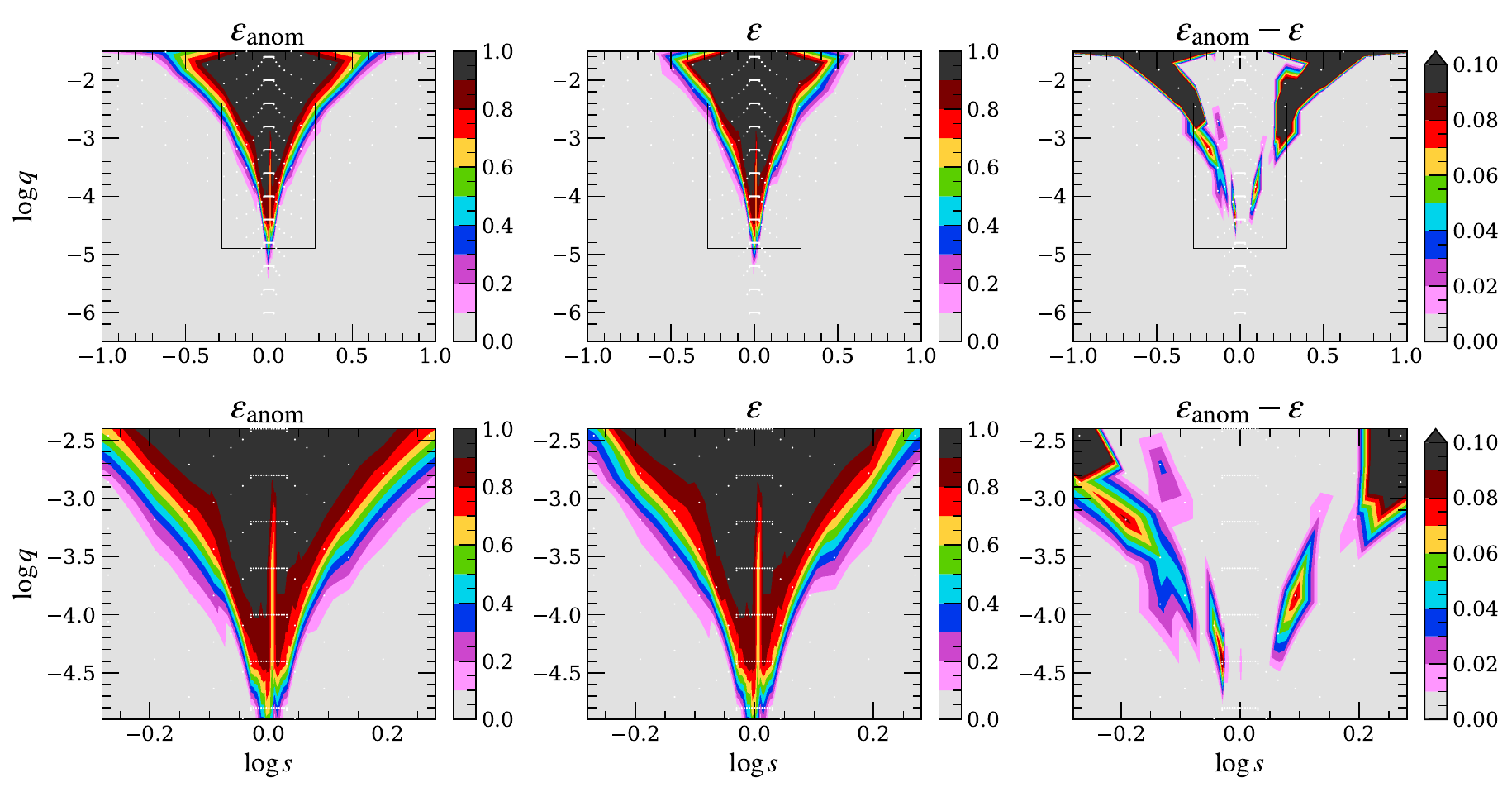}
    \caption{The $\alpha$-integrated sensitivity maps $\eanom$ (without accounting for 2L1S degeneracy), $\e$ (with 2L1S degeneracy included), and their difference for the DSHM group. The white dots indicate the actual sampled points in the simulation, and the full map is linearly interpolated in the $(k, h)$ space. The upper panels show the complete maps, and the lower ones provide zoomed-in views of the boxed region, where $k$ and $h$ are defined in eqs. \ref{eq:kdef} and \ref{eq:hdef}.}
    \label{fig:sens_DSHM}
\end{figure*}

We now quantify the impact of degeneracy on sensitivity from a statistical perspective. 
As described in Section~\ref{sec:method}, we obtain both the sensitivity without accounting for 2L1S degeneracy, $\eanom(\log s, \log q, \alpha)$, and the sensitivity after considering such degeneracy, $\e(\log s, \log q, \alpha)$, as functions of $\log s$, $\log q$, and $\alpha$. 
Because $\alpha$ is a geometric parameter unrelated to physical properties and is randomly distributed in actual events, we first integrate over $\alpha$:
\begin{align}
     \eanom(\log s,\log q) & = \int{\eanom(\log s,\log q,\alpha)\dd\alpha}  \\
     & \approx \frac{1}{N_{\alpha}}\sum_{i=1}^{N_{\alpha}}{\eanom(\log s,\log q,\alpha_i)},
\end{align}
and similarly
\begin{align}
     \e(\log s,\log q) & = \int{\e(\log s,\log q,\alpha)\dd\alpha}  \\
     & \approx \frac{1}{N_{\alpha}}\sum_{i=1}^{N_{\alpha}}{\e(\log s,\log q,\alpha_i)},
\end{align}
where $N_{\alpha}$ is the number of simulated $\alpha$ values, $N_{\alpha}=60$ for the GSHM, DSHM, and DSEM groups, and $N_{\alpha}=180$ for the GSLM and DSLM groups. 

Figure \ref{fig:sens_DSHM} shows the $\alpha$-integrated sensitivity maps $\eanom(\log s, \log q)$, $\e(\log s, \log q)$, and their difference for the DSHM group as an example. The white dots represent the actual simulation grids, and the full map is obtained by linear interpolation in the $(k, h)$ space.
Several notable features are visible in Figure~\ref{fig:sens_DSHM}.
First, the $(k, h)$ grids successfully capture the decrease in sensitivity near the very narrow region around $\log s \sim 0$. 
This is consistent with the prediction by \citet{Chung2009_ResonantCaustic} that the perturbation from a resonant caustic is relatively weak in high-magnification events. 
\HL{Such sensitivity suppression can be ameliorated if higher-cadence follow-up observations are taken.
In our case, the suppression} region roughly spans $0<\log s<0.01$ and $-4.4<\log q<-3.0$, which is so small that it is often insufficiently sampled by conventional evenly spaced $\log s$ grids.
Second, sensitivity also decreases near the upper boundary of the sampled region around $\log q \sim -1.5$, for both $\eanom$ and $\e$. This is an artifact caused by insufficient sampling near the boundary and limitations of the interpolation method.
Third, the difference between $\eanom$ and $\e$ is mostly located near the edge of the sensitive region. This occurs because signals in these transition zones (where $\eanom$ approaches 0) are generally subtle and more easily affected by degeneracies.
Last, the sensitivity decrease is significantly asymmetric with respect to $\log s=0$. 
This asymmetry arises from two factors: the inherent asymmetry of planetary caustic for $\log s<0$ and $\log s>0$, as illustrated in Figure~\ref{fig:lc_exp2}, and the intrinsic, slight differences between ``close'' and ``wide'' central caustics and their magnification patterns \citep[\HL{for example}, ][]{KemingZhang2021_OffsetDegeneracy,Zhang2022_OffsetDegeneracy}. The latter is also reflected in the first feature, where the resonant sensitivity drop is not centered at $\log s = 0$ but rather near $\log s \sim 0.006$.
More specifically, the region $-0.2\lesssim\log s\lesssim-0.1$ and $-4.0\lesssim\log q\lesssim-2.5$ is dominated by the planetary caustic ``close-wide'' degeneracy, while the other regions are mainly affected by the ``central-resonant'' degeneracy.

\begin{figure*}
    \centering
    \includegraphics[width=0.7\columnwidth]{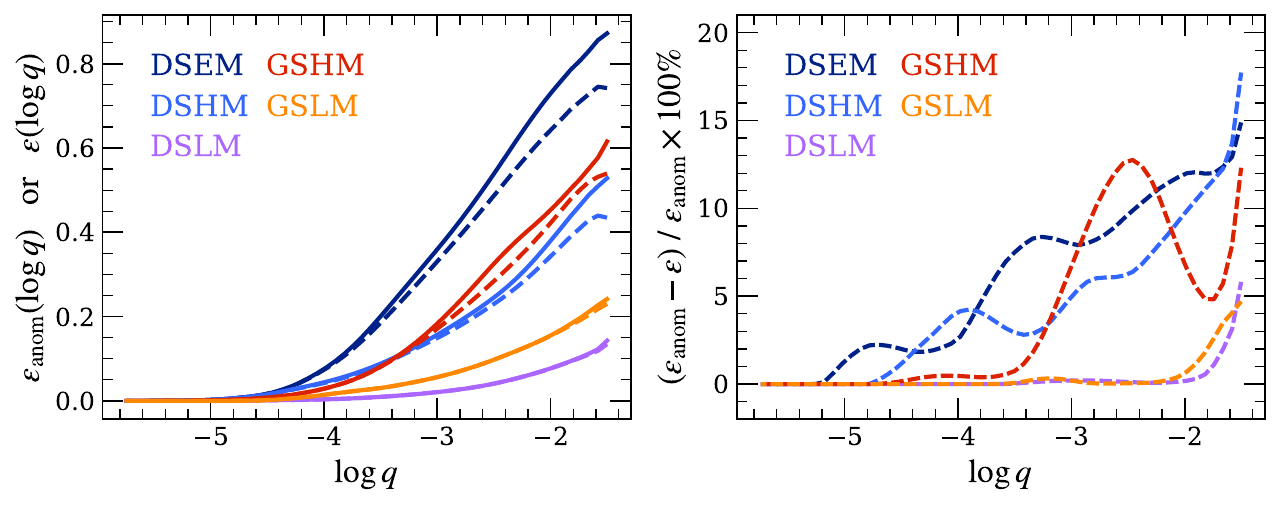}
    \caption{Left: The $(\alpha, \log s)$-integrated sensitivities $\eanom$ (solid lines) and $\e$ (dashed lines) as a function of $\log q$ for all event groups. Right: The relative decrease in sensitivity after accounting for $\edeg$ for all groups.}
    \label{fig:sens_logq}
\end{figure*}

To investigate the influence of degeneracy as a function of $\log q$, we further integrate $\log s$ for both $\eanom$ and $\e$, 
\begin{align}
    \eanom(\log q) &= \int{\eanom(\log s,\log q)\dd\log s}/\int{\dd\log s},\\
    \e(\log q) &= \int{\e(\log s,\log q)\dd\log s}/\int{\dd\log s}.
\end{align}
Because the sampling points are non-uniform in $(\log s, \log q)$ space, we first linearly interpolate both $\eanom$ and $\e$ in the $(k, h)$ space, and numerically compute the integrals over $(\log s, \log q)$. The absolute value of the integrated sensitivity depends on the integration region in $\log s$ \citep[\HL{for example}, ][]{Suzuki2016}. However, the relative difference between $\eanom$ and $\e$ is not sensitive to the integration range. We perform the integration over $-1.0 \leq \log s \leq 1.0$.

Figure \ref{fig:sens_logq} shows the $(\alpha, \log s)$-integrated sensitivities $\eanom$ and $\e$ as functions of $\log q$, along with their relative differences, for each event group. Overall, the fractional decrease in sensitivity increases toward larger $\log q$, though each group has distinct fine-scale features.

The two low-magnification groups, DSLM and GSLM, show only mild sensitivity decreases except near $\log q=-1.5$. 
In contrast, the other three groups exhibit clear peak and dip features across $\log q$.
For the DSEM group, the peak at $\log q\sim-4.8$ is dominated by the ``central-resonant'' caustic degeneracy, as illustrated in Figure \ref{fig:lc_exp3}, and the other peaks arise from a combination of the ``central-caustic'' and ``planet-binary'' degeneracies. This occurs because, under extreme magnification, anomaly signals are primarily produced by central caustic interactions.
In the high-magnification groups, DSHM and GSHM, planetary caustics dominate the sensitivity loss at lower mass ratios ($\log q\lesssim-3.5$), as in the example shown in Figure \ref{fig:lc_exp2}. This is because the central caustic becomes relatively small compared to the impact parameter $u_0 = 0.02$ \citep{Chung2005_CentralCaustic}.
The peaks at $\log q\gtrsim-3.5$ are dominated by the ``planet-binary'' degeneracy, as \HL{illustated} in Figure \ref{fig:lc_exp1}.
\HL{However, the peaks in the right panel of Fig. \ref{fig:sens_logq} may be due to our limited choices of the source size $\rho$ and impact parameter $u_0$ as shown in Table~\ref{tab:simu_summary}, and can be largely smoothed when combining a large number of events with various parameters, as will be discussed in Section~\ref{sec:dis_mf}.}

\begin{figure*}
    \centering
    \includegraphics[width=1.0\columnwidth]{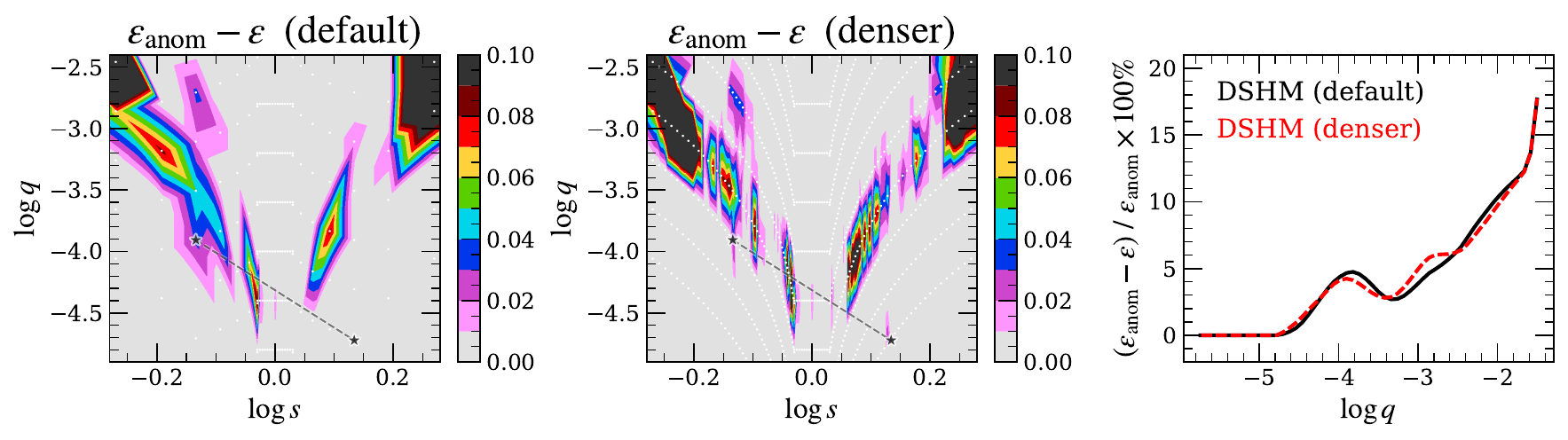}
    \caption{Left and middle: The $\alpha$-integrated sensitivity difference maps for the DSHM group under default (left) and denser (middle) sampling grids. The ``close–wide'' degenerate pair from the event in Figure \ref{fig:lc_exp2} is marked with star symbols connected by a dashed line. Right: Relative decrease in $(\alpha, \log s)$-integrated sensitivity as a function of $\log q$ for the default (solid black) and denser (dashed red) sampling grids.}
    \label{fig:sens_diff_dense}
\end{figure*}

We notice that the current sampling in $(\log s, \log q)$ space may be insufficient to capture all degeneracy cases. 
For example, in the simulated event shown in Figure \ref{fig:lc_exp2}, the true solution is located at $(\log s, \log q, \alpha) = (-0.134, -3.907, 0.0^\circ)$, with a corresponding degenerate ``wide'' solution at $(\log s, \log q, \alpha) = (0.134, -4.723, 181.7^\circ)$. However, due to sparse sampling in $\alpha$, this wide solution was not initially detected ($\eanom = 0$). Similarly, some other degeneracy pairs are missed because of the sparse sampling in $(\log s, \log q)$ space.
Therefore, to evaluate the potential impact of sampling density, we repeated the analysis with a denser sampling in $(\log s, \log q, \alpha)$ space for the DSHM group within the zoomed-in region highlighted in Figure~\ref{fig:sens_DSHM}.
The refined grids successfully recovered the previously missed solutions.
While some fine-scale features differ between the two sampling schemes, the overall morphology of the sensitivity decrease remains consistent. 
Thus, our conclusion that the 2L1S degeneracies reduce planetary sensitivity by approximately $10\%$ remains valid.

In summary, our simulations indicate that 2L1S degeneracies decrease microlensing planetary sensitivity at the level of $5\%$ to $10\%$ level, and the effect becomes stronger when $\log q$ increases.

\section{Discussion}\label{sec:dis}

\subsection{Estimate the Impact on Mass-ratio Functions}
\label{sec:dis_mf}

In Section \ref{sec:result}, we derive the effects of degeneracies across five representative groups of simulated events. We now extend these results to a realistic statistical sample and estimate the overall impact of degeneracy on \HL{planet} sensitivity.

We use the KMTNet sample as an illustrative case. We count the microlensing events found by KMTNet in the 2018 and 2019 seasons, which is the same sample used by \citet{Zang2025Science_KMT_MassRatioFunction} for sensitivity calculations. 
We exclude non-microlensing events and binary-lens events with $\log q > -1.5$, resulting in a total of 5,333 events. They are categorized into the five groups (GSLM, GSHM, DSLM, DSHM, DSEM) listed in Table \ref{tab:setup} according to the following criteria:
\HL{
\begin{align}
    {\rm GSLM}&:~(u_0>0.26)~{\rm and}~(I_{\rm S}-A_I\leq17), \\
    {\rm GSHM}&:~(u_0\leq0.26)~{\rm and}~(I_{\rm S}-A_I\leq17), \\
    {\rm DSLM}&:~(u_0>0.10)~{\rm and}~(I_{\rm S}-A_I>17), \\
    {\rm DSHM}&:~(0.012\leq u_0<0.10)~{\rm and}~(I_{\rm S}-A_I>17), \\
    {\rm DSEM}&:~(u_0\leq0.012)~{\rm and}~(I_{\rm S}-A_I>17),
\end{align}
}
where $A_I$ is the $I$-band \HL{extinction}. \HL{The $u_0$ classification boundaries are set based on the cumulative distribution of the observed sample for giant-source and dwarf-source, respectively. Specifically, for each boundary between two simulated $u_0$ values (from Table~\ref{tab:setup}), we set the corresponding observational boundary at the $u_0$ value whose cumulative fraction is the midpoint of those of the simulated values. This ensures a consistent proportional mapping between the simulation parameters and the observed sample properties.}
\begin{equation}
    (n_{\rm GSLM}: n_{\rm GSHM}: n_{\rm DSLM}: n_{\rm DSHM}: n_{\rm DSEM}) = \HL{(234 : 1355 : 2432 : 352 : 960)}.
\end{equation}
We also account for the multi-cadence observing strategy of the KMTNet survey. Approximately $(12, 28, 44, 12)~{\rm deg^2}$ of the sky are observed with cadences of $\Gamma=(4, 1, 0.4, 0.2)~{\rm hr^{-1}}$, respectively. The number of events in each cadence group is
\begin{equation}
    (n_{\rm 4\,hr^{-1}} : n_{\rm 1\,hr^{-1}} : n_{\rm 0.4\,hr^{-1}} : n_{\rm 0.2\,hr^{-1}}) = (1822 : 1913 : 1348 : 250).
\end{equation}
Because our simulations assume a cadence of $\Gamma = 4~\mathrm{hr}^{-1}$, we rescale the $\chi^2$ values for other cadences by factors of $\frac{1}{4}$, $\frac{1}{10}$, and $\frac{1}{20}$ for $\Gamma = 1$, $0.4$, and $0.2~\mathrm{hr}^{-1}$, respectively, reflecting the proportionality $\chi^2 \propto N_{\mathrm{data}} \propto \Gamma \tE$, where $N_{\mathrm{data}}$ is the number of data points. We ignore the $\tE$ factor because we already use the median value as the default setting.
Specifically, for the $\Gamma$ factor, we multiply the measured $\chi^2$ by these factors and reapply the anomaly detection and model comparison procedures described in Section \ref{sec:method} to reassign $\eanom$ and $\edeg$. \HL{This approach neglects any potential new minima that might arise as a result of the decreased coverage.}

After accounting for event categories and cadence effects, we combine the sensitivities of individual events to obtain the overall survey sensitivity,
\begin{equation}
    S_{\rm anom}(\log q) = \sum_{i}^{N_{\rm event}}{\varepsilon_{{\rm anom},i}}(\log q)
\end{equation}
and 
\begin{equation}
    S(\log q) = \sum_{i}^{N_{\rm event}}{\varepsilon_{i}}(\log q),
\end{equation}
where $S$ and $S_{\rm anom}$ represent the sample sensitivity with and without considering the 2L1S degeneracy effect, respectively.

\begin{figure*}[htb]
    \centering
    \includegraphics[width=0.4\columnwidth]{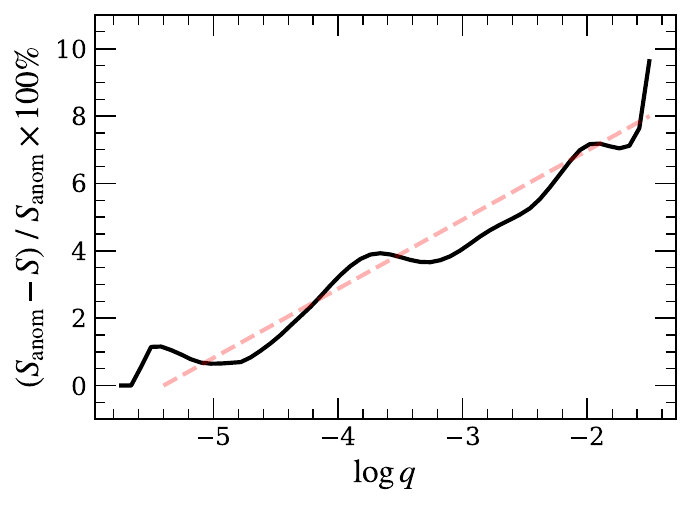}
    \caption{Estimated decrease in the sample planetary sensitivity of a microlensing survey (\HL{for example}, KMTNet, \citealt{Kim2016_KMT}) due to 2L1S degeneracy, as a function of the planet–host mass ratio $q$. }
    \label{fig:sens_sample}
\end{figure*}

Figure \ref{fig:sens_sample} shows the relative difference in sample sensitivity before and after accounting for degeneracies. The bump at $\log q\sim-5.4$ is probably an artifact due to the limited sampling at the small $q$ end. Besides that, we find that the effect of degeneracy generally increases smoothly with $\log q$  (as indicated by the red dashed line), ranging from approximately $0\%$ at $\log q \simeq -5$ to about $10\%$ at $\log q \simeq -1.5$. The peaks shown in Figure \ref{fig:sens_logq} are smoothed after combining a large number of events from different groups and cadences. 
We also note that the overall shape of the sensitivity decrease is not highly sensitive to the exact event counts in each group, though it does influence the slope of the curve.

Because the sensitivity reduction due to 2L1S degeneracy depends on $\log q$, it introduces a bias in measurements of the mass-ratio function. 
\HL{For example, the slope of the mass-ratio function \citep[e.g.,][]{Suzuki2016, Zang2025Science_KMT_MassRatioFunction} would become less steep when the effect is included.}
\HL{In the survey-follow-up hybrid \citet{Suzuki2016} sample, the higher fraction of high-magnification events and the higher effective cadences near the peak influence the degeneracy-induced sensitivity decrease in opposite directions, therefore, it is not clear whether the net effect is stronger or weaker.}
Nevertheless, the overall effect is relatively modest ($\lesssim 10\%$) and may not be significant in current microlensing statistical samples\HL{, which} contain $<100$ planets. However, the effect will become considerably more important in future surveys such as those conducted by the \emph{Roman} \citep{Spergel2015_WFIRST, MatthewWFIRSTI} and \emph{Earth 2.0} (ET) \citep{Gould2021_FFP_ET,ET_WhitePaperv1} missions, which are expected to detect $\mathcal{O}(10^3)$ planets.

\subsection{Future Application \HL{to} Real Events}
\label{dis:future}

We estimated the degeneracy effect on the mass-ratio function in Section \ref{sec:dis_mf}. However, these estimates are not sufficiently accurate to be directly applied in statistical studies, as real events are abstracted into one of the five groups defined in Table \ref{tab:setup}, with strictly uniform time sampling and an idealized signal-to-noise ratio curve that does not account for real observational gaps or systematic errors.

A rigorous approach to incorporating this effect for real events would involve using the actual time sampling and photometric errors of each event and applying the full procedure outlined in Section \ref{sec:method}.
However, the computational cost is a major challenge.
The dominant part of the time cost is the global parameter grid search. Although we employed a pre-generated magnification map method (Zhang et al., in prep) to improve efficiency, the typical time required for the grid search per light curve is approximately
\begin{equation}
    T =1.21\times10^{-7} (N_{\rm data}N_{\rm grid}N_{\alpha})~{\rm CPU\cdot hr},
\end{equation}
where $N_{\rm data}$ is the number of data points in the light curve, $N_{\rm grid}$ is the number of the $(\log s, \log q)$ grids (and sometimes $\rho$ grids, if needed), and $N_{\alpha}$ is the number of initial $\alpha$ values. 
In our simulations, $(N_{\rm data}, N_{\rm grid}, N_{\alpha}) = (8640, 533, 8)$, resulting in a \HL{grid-search} cost of about $4.5~{\rm CPU\cdot hr}$ per light curve.
To sample the sensitivity of one event across the full 2L1S parameter space, approximately $10^4$ simulated light curves are required to undergo grid searches, totaling around \HL{$4.5 \times 10^4~{\rm CPU\cdot hr}$} per event.
Moreover, a statistical sample often consists of $>10^3$ events. For example, \citet{Suzuki2016} used 1,474 events and \citet{Zang2025Science_KMT_MassRatioFunction} used 3,657 events. 
Thus, the total computational cost for a full statistical sample would be on the order of \HL{$\mathcal{O}(10^8)~{\rm CPU\cdot hr}$}.
\HL{Porting the grid search to GPUs will principally reduce the computational cost from $\mathcal{O}(10^8)~{\rm CPU\cdot hr}$ to $\mathcal{O}(10^5)~{\rm GPU\cdot hr}$, because each GPU chip offers a floating-point performance and memory bandwidth comparable to $\gtrsim10^3$ CPU cores for such parallelizable tasks. Nevertheless, this still represents a substantial computational effort.}
Additionally, as shown in Section \ref{sec:example:central-resonant}, the $(k, h)$ grid used here still failed to resolve two very close resonant local minima, indicating that the parameter space remains undersampled.
Furthermore, the $(k, h)$ parameterization is also inefficient for sampling planetary-caustic anomaly events.
A comprehensive sampling of the parameter space would require a combination of $(k, h)$ and $(\log s, \log q)$ grids, demanding even greater computational resources.

Therefore, applying the current rigorous CPU-based algorithm to incorporate 2L1S degeneracy effects across an entire statistical sample is computationally unrealistic.
In the future, several feasible paths are worth exploring.
The first is to follow the approach used in Section \ref{sec:dis_mf}, which abstracts a small set of typical events and extrapolates the results to the full sample.
The second is to perform rigorous sensitivity calculations only on a homogeneous subset of events to represent the entire sample, or inject signals with random parameters rather than using fixed grids. Such methods are adopted by one of the early microlensing statistics \citep{Cassan2012Nature}, and are commonly used in the transit planet statistics \citep[e.g., ][]{Fressin2013_KeplerOccurrenceRate,Christiansen2015_KeplerDetectionEfficiency,Gan2023_MdwarfTransit}.
However, the accuracy of these approximate methods requires further validation.
\HL{The third path is, as already mentioned above, to explore using GPU clusters to accelerate statistical analyses.}
The fourth path and likely the most promising way to balance accuracy and efficiency, is to develop machine learning methods to explore the full parameter space for candidate solutions \citep[e.g., ][]{ZhaoZhu2022_MAGIC,Zhang2021_RomanAIModel}. 
A key challenge on this path is ensuring the robustness of machine learning models when applied to real observational data.

Beyond \HL{the} sensitivity calculation, \HL{for} real events, it is sometimes possible to break 2L1S degeneracies. 
As noted in Section \ref{sec:intro}, degeneracies arise from both intrinsic similarities among magnification patterns and the fact that the light curve represents only a one-dimensional trace through the magnification map. 
Therefore, a straightforward way to resolve degeneracies is to obtain additional samples of the magnification map.
Improved sampling can be achieved through the parallax effect \citep{Gould1992, Gould2000_Formalism, Gouldpies2004}, orbital motion of the lens or source system \citep[``xallarap'' effect][]{Griest1992_1L2SEffect, Han_Gould_1997_1L2Sxarallap}, or co-observing with a space telescope \citep[e.g.,][]{Gould2021_FFP_ET}.
However, incorporating degeneracy effects into the sensitivity calculation for such events remains an unresolved and complex problem, which lies beyond the scope of this study.

\acknowledgments
H.Y. acknowledge support by the China Postdoctoral Science Foundation (No. 2024M762938).
Y.S., H.Y., J.Z., Q.Q., W.Z., and S.M. acknowledge support by the National Natural Science Foundation of China (Grant No. 12133005). 
W.Z. acknowledges the support from the Harvard-Smithsonian Center for Astrophysics through the CfA Fellowship. 
This work is part of the ET space mission which is funded by the China's Space Origins Exploration Program.
The authors acknowledge the High-performance Computing center at Westlake University and the Tsinghua Astrophysics High-Performance Computing platform at Tsinghua University for providing computational and data storage resources that have contributed to the research results reported within this paper. 
The authors thank R. Poleski, D. Bennett, and the anonymous reviewers for their helpful comments.

\bibliography{Yang.bib}

\end{CJK*}
\end{document}